\documentclass[12pt,twoside,A4,reqno]{amsart}
\usepackage{epsfig}

\calclayout

\def\({\left(}
       \def\){\right)}
       \def\[{\left[}
       \def\]{\right]}
    \def\<{\langle}
    \def\>{\rangle}

\newcommand{\bea}{\begin{eqnarray}}
\newcommand{\eea}{\end{eqnarray}}
\begin{document}
\title
  {Open-Loop Control in Quantum Optics:\\ Two-Level Atom in Modulated Optical Field}
\maketitle
\begin{center}\author{Saifullah$^1$,\, Sergei Borisenok $^2$}\\
\vspace{.5cm} {$^1$Abdus Salam School of Mathematical Sciences,\\ Government College University,\\
35 C - II, Gulberg III, Lahore, Pakistan\\
\vspace{0.2cm}
Email: saifullahkhalid75@yahoo.com
\\
\vspace{0.5cm}}

$^{2}$ Deptt. of Physics, Herzen State Pedagogical University\\
48 Moika River, 191186 St. Petersburg, Russia\\
\vspace{0.2cm}
Email: sebori@mail.ru
\end{center}
\begin{abstract}
The methods of mathematical control theory are widely used in the
modern physics, but still they are less popular in quantum science.
We will discuss the aspects of control theory, which are the most
useful in applications to the real problems of quantum optics. We
apply this technique to control the behavior of the two-level
quantum particles (atoms) in the modulated external optical field in
the frame of the so called "semi classical model", where quantum
two-level atomic system (all other levels are neglected) interacts
with classical electromagnetic field. In this paper we propose a
simple model of feedforward (open loop) control for the quantum
particle system, which is a basement for further investigation of
two-level quantum particle in the external one-dimensional optical
field.
\end{abstract}
\maketitle

\pagestyle{myheadings} \markboth{\centerline {\scriptsize Saifullah,\, Sergei Borisenok}} {\centerline {\scriptsize  Open-Loop Control in Quantum Optics: Two-Level Atom in Modulated Optical Field}}
\section{Introduction}
\noindent In this paper we present a semiclassical theory of the
interaction of a single two-level atom with a single mode of the
field in which the atom is treated as a quantum two-level system and
the field is
treated classically.\\
We will use a simple physical model in order to define the goals,
terminology, and methodology of applied control theory.\\
We propose a simple model of feedforward (open-loop) control for
two-level atom in the external one-dimensional optical
field.\\
We use the so-called "semi-classical model" of the atom-field
interaction that describes a single quantum two-level atomic system
(all other levels are neglected) with classical electromagnetic
field. We use the standard notation following \cite{Scully}, but in
our model the optical field plays the role of a control signal
$u(t)$ for open-loop (feedforward) control scheme \cite{Fradkov}.
Recently other authors studied the control of two-level atoms in the
frame of open-loop ideology when the controlling field was known a
priori. It allowed to get the different forms of atomic energy
spectra, producing $\pi$- and $\frac{\pi}{2}$- pulses \cite{Imoto},
taking special non-constant shapes of external field \cite{Di
Piazza} etc.
\section{Interaction of two-level atom with open loop control optical field}
\noindent We consider the quantum two-level atomic system in the
classical optical field $E(t)$. Let $|a\rangle$  and $|b\rangle$
represent the upper and lower level states of the atom, i.e., they
are eigenstates of the unperturbed part of the Hamiltonian
$\hat{H_0}$ with the eigenvalues: $\hat
H_{0}|a\rangle=\hbar\omega_a|a\rangle$
\,\,\,\,\,and\,\,\,\,\, $\hat H_{0}|b\rangle=\hbar\omega_b|b\rangle$.\\
The equations of motion for the density matrix elements are given by
\cite{Fradkov}:
\begin{eqnarray}
\nonumber &&\dot{\rho_{aa}}=-\gamma_a\rho_{aa}+\frac{\iota
E}{\hbar}\Big(P_{ab}\rho_{ba}e^{\iota\omega
t}-P^{*}_{ab}\rho_{ab}e^{-\iota\omega t}\Big) \ ;\\
&&\dot{\rho_{bb}}=-\gamma_b\rho_{bb}-\frac{\iota
E}{\hbar}\Big(P_{ab}\rho_{ba}e^{\iota\omega
t}-P^{*}_{ab}\rho_{ab}e^{-\iota\omega t}\Big) \ ; \\
\nonumber &&\dot{\rho_{ab}}=-\gamma_{ab}\rho_{ab}-\frac{\iota
E}{\hbar}P_{ab}\Big(\rho_{aa}-\rho_{bb}\Big)e^{\iota\omega t} \ ,
\end{eqnarray}
where $\rho_{ba}=\rho^{*}_{ab}$ \, ;\,\, $P_{ab}$  is the matrix
element of the electric dipole moment, $\gamma_a$ and $\gamma_b$ are
the decay constants,
$\gamma_{ab}=\frac{\gamma_a+\gamma_b}{2}+\gamma_{ph}$ is a decay
rate including elastic collisions between atoms, and
$\omega=\omega_a - \omega_b$ is the atomic transition frequency.\\
Let's denote $P_{ab}=|P_{ab}|e^{\iota\varphi}$  and
\begin{eqnarray}
\nonumber &&\rho_{+}=\rho_{ba}e^{\iota(\omega t+\varphi)}+
\rho_{ab}e^{-\iota(\omega t+\varphi)} \ ;\\
&&\rho_{-}=\iota\Big[\rho_{ba}e^{\iota(\omega
t+\varphi)}-\rho_{ab}e^{-\iota(\omega t+\varphi)}\Big].
\end{eqnarray}
Using (2) we can re-write the system (1) in the real form:
\begin{eqnarray}
\nonumber &&\dot{\rho_{aa}}=-\gamma_a\rho_{aa}+\frac{|P_{ab}|
E}{\hbar}\rho_{-} \ ;\\
\nonumber &&\dot{\rho_{bb}}=-\gamma_b\rho_{bb}-\frac{|P_{ab}|
E}{\hbar}\rho_{-} \ ; \\
&&\dot{\rho_{+}}=-\gamma_{ab}\rho_{+}+\omega\rho_{-} \ ; \\
\nonumber
&&\dot{\rho_{-}}=-\gamma_{ab}\rho_{-}-\omega\rho_{+}-\frac{2|P_{ab}|
E}{\hbar}\Big(\rho_{aa}-\rho_{bb}\Big).
\end{eqnarray}
For further calculation we put  $\gamma_a=\gamma_b=\gamma$.\\ Then
\begin{equation}
\Big(\rho_{aa}+\rho_{bb}\Big)(t)=e^{-\gamma
t}\Big(\rho_{aa}+\rho_{bb}\Big)(0).
\end{equation}
The first two equations of system (3) can be combined together.\\
We put:
\begin{eqnarray}
\nonumber &&\rho_{aa}-\rho_{bb}\equiv e^{-\gamma t}x(t) \ ;\\
&&\rho_{+}\equiv e^{-\gamma t}y(t) \ ; \\
\nonumber &&\rho_{-}\equiv e^{-\gamma t}z(t).
\end{eqnarray}
Substituting (5) in (3) we eliminate the decay $\gamma$-containing
terms.\\ Finally, re-scaling the time by $\omega$: $\tau =\omega t$,
and denoting the dimensionless control signal by $u(t)\equiv \frac{
2|P_{ab}|E(t)}{\hbar\omega}$ and $\epsilon =
\frac{\gamma_{ph}}{\omega}$, we get the simplified system:
\begin{eqnarray}
\nonumber &&\dot{x}=u(t) z \ ;\\
&&\dot{y}=-\epsilon y + z \ ; \\
\nonumber &&\dot{z}=-\epsilon z - y - u(t) x.
\end{eqnarray}
Here the dot means the derivative with respect to the new
dimensionless time  $\tau$.\\ We remind that $x\in[-1, 1]$, since
$\rho_{aa}-\rho_{bb}\in[-1, 1]$, and $\rho_{aa}-\rho_{bb}
\rightarrow 0$ as $t \rightarrow \infty$.
\section{The case of conservative system}
\noindent We mention here the special (particular) case of
$\gamma_{ph}=0$ and thus, $\epsilon=0$.\\ Then the system (6) has
the integral of motion:
\begin{equation}
x^2 + y^2 + z^2 \equiv r^2 = \hbox{const}.
\end{equation}
Now in the spherical coordinates:
\begin{eqnarray}
\nonumber &&x(\tau) = r\cos\alpha(\tau)\sin\beta(\tau) \ ;\\
&&y(\tau) = r\sin\alpha(\tau)\sin\beta(\tau) \ ; \\
\nonumber &&z(\tau) = r\cos\beta(\tau).
\end{eqnarray}
Then
\begin{eqnarray}
\nonumber &&\dot{\alpha}= \cot\beta\Big(\cos\alpha - u\sin\alpha\Big) ;\\
&&\dot{\beta}= \sin\alpha + u \cos\alpha.
\end{eqnarray}
If  $x(0) = -1$,  $y(0) = z(0) = 0$, then $\alpha(0) = 0$  and
$\beta(0) = \frac{-\pi}{2}$.
\section{Numerical solution}
\noindent To solve (6) numerically we will use Maple.\\ Let's apply
the initial conditions  $$x(0) = -1\, ,\,\,\,\,\,  y(0) = z(0) = 0
,$$ corresponding
to the ground level of the atom.\\
Here we put $\epsilon = 0.01$ and $u(\tau) = \cos\tau$.\\ The result
is presented on Fig.1 (left).\\ Another case corresponds to the
increasing signal $u(\tau) = e^{-\tau}\cos\tau$.\\ The result is
presented on Fig.1 (right).\\
\begin{figure}[h]
 \begin{center}
  \includegraphics[width=11 cm]{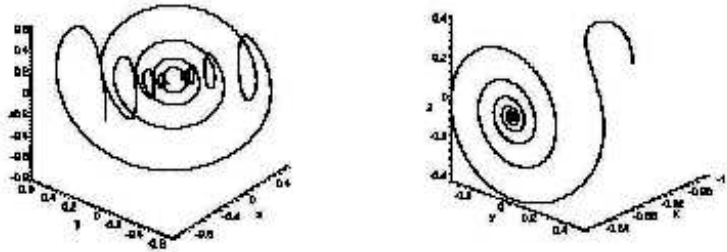}\\
  \caption{Phase portraits of the dynamical system (6) with $\epsilon = 0.01$ and the
  control signal $u(\tau)=\cos\tau$ (left) and $u(\tau)=e^{-\tau}\cos\tau$ (right).}\label{}
  \end{center}
  \end{figure}\\
\noindent From Fig.1 we can conclude that the shape of the control
field (i.e. the optical field $E$) influences drastically on the
present state of two-level atoms.
\section{Conclusion}
\noindent In this paper we studied the techniques to control the
behavior of the two-level quantum particles by the modulated
external optical field.\\ We conclude that the semi-classical model
of interaction between two-level quantum particle system and
classical optical field can be successfully applied to describe the
control process of the particle energy stabilization.

\end{document}